\documentclass{fairmeta}

\usepackage{amsmath, amssymb, amsthm}
\usepackage{tikz}
\usetikzlibrary{positioning, arrows.meta}
\definecolor{metablue}{HTML}{0064E0}
\definecolor{metaorange}{HTML}{DC6414}

\usepackage{fancyhdr}
\IfFileExists{gitrev.tex}{\newcommand{\gitstamp}{}
}{}
\providecommand{\gitstamp}{version unknown}
\fancypagestyle{plain}{%
  \fancyhf{}%
  \fancyfoot[C]{\thepage}%
  \fancyfoot[R]{\footnotesize\textcolor{gray}{\gitstamp}}%
}

\newtheorem{proposition}{Proposition}
\newtheorem{lemma}{Lemma}

\newtheorem{remark}{Remark}

\title{Interpretable and Validatable Uplift Modeling}

\author[1,*]{Bob Wilson}
\affiliation[1]{Marketing Data Science at Meta}
\contribution[*]{Work done at Meta}
\date{October 2026}
\correspondence{Bob Wilson at \email{rwilson4@meta.com}}

\abstract{Uplift models estimate treatment effects for individuals,
  but we do not observe individual treatment effects, so the
  predictions are hard to interpret and validate. A decision-maker
  often wants trustworthy claims about groups where the treatment is
  beneficial or harmful, and to know when a treatment benefits some
  people more than others. Our procedure provides such a set of
  claims, certified simultaneously at a specified familywise error
  rate (FWER). We summarize an uplift model by a shallow tree fit to
  its predictions. Each node of the tree carries two hypotheses, one
  about the average treatment effect in the node and one about the
  difference in effects across the node's split; the two kinds are
  interwoven into a single hierarchy. Splitting the data enables
  honest estimation in the evaluation set. It lets us choose the
  \emph{direction} of each one-sided test from the training half,
  allowing us to certify the sign of treatment effects in
  subpopulations, and to compare effects between groups. We
  demonstrate our procedure offers strong control of the FWER.}

\begin{document}

\maketitle

\section{Introduction}
\label{sec:intro}

Uplift modeling uses observed covariates, \(X\), to estimate the
impact of a treatment on each individual person:
\begin{displaymath}
    \tau(x) := \mathbf{E}[ Y(1) - Y(0) \, | \, X=x],
\end{displaymath}
where \( (Y(0), Y(1) \) are potential outcomes.
\citet{gutierrez-2017-uplift} offer an introduction to the vast
literature on uplift modeling.

Since individual treatment effects are unobservable, it can be
difficult to validate an uplift model. Many researchers use random
forests \citep{wager-2018-uplift} or boosted trees
\citep{powers-2018-uplift} as model families, which are hard to
interpret.

Previous work uses Qini or lift curves to summarize how well the model
orders individuals by effect. The GATES of
\citet{chernozhukov-2018-generic} report average effects within
quantile bins of the predicted score. The RATE of
\citet{yadlowsky-2025-rate} condenses targeting value into a single
statistic with a confidence interval. These methods do not deliver
insights about how the treatment performs in meaningful
subpopulations. It is valuable to be able to say things like,
\emph{the treatment benefits men more so than women, but overall it is
beneficial for both groups,} and to attach some claim of evidentiary
strength.

Our method allows researchers to generate such claims, and to certify
them as statistically significant. We split the data into training and
evaluation sets. We fit the uplift model on the training set, then
summarize it with a shallow regression tree. Turning to the evaluation
set, we attach to each node two hypotheses: one about the average
effect in the node, one about the difference in effects across the
node's split. We interweave the two kinds into a single hierarchy
tested by a procedure controlling the familywise error rate, producing
a set of trustworthy claims about meaningful subpopulations. More than
reporting a single average treatment effect for the entire study
population, our method offers rich insights about how treatment
effects vary across the population, providing greater confidence or
suitable caution to the decision maker.

The procedure rests on four components. The first splits the data to
facilitate honest estimation. The second interweaves \emph{effect} and
\emph{heterogeneity} hypotheses (\S\ref{sec:interweaving}), which lets
the procedure consider treatment effects \emph{within} subpopulations
and differences in treatment effects \emph{across} subpopulations. The
third adapts the direction of each test to the training half of the
data (\S\ref{sec:adaptive}): one-sided tests certify the treatment as
beneficial or harmful in particular sub-groups, and more beneficial in
some groups than others. The fourth sets the levels of individual
tests to guarantee strong control of the FWER. None of these
contributions are novel individually, but we have not seen them
applied in combination in the context of uplift modeling.

\S\ref{sec:interpretable} sets up the tree summary and the data split.
\S\ref{sec:interweaving} defines the hypotheses and their
interweaving. \S\ref{sec:adaptive} covers adaptive direction
selection. \S\ref{sec:fwer} handles multiplicity. \S\ref{sec:bounds}
offers a confidence bound as a consolation prize when we fail to
reject a hypothesis. \S\ref{sec:simulation} presents the simulation
study, and \S\ref{sec:summary} concludes.

\section{Interpretable Uplift Modeling}
\label{sec:interpretable}

Researchers often use shallow decision trees to aid interpretability
of otherwise inscrutable models, including for uplift modeling. The
Virtual Twins procedure of \citet{foster-2011-virtual-twins} fits a
tree to individual effect estimates from a random forest. The
aggregation trees of \citet{difrancesco-2026-aggregation-trees} build
a nested sequence of subgroups from estimated conditional effects.
\citet{wu-2022-interpretable-personalized} distill personalized
experimentation results for interpretability at scale.
\citet{huang-2025-distilling-hte} pursue stable subgroup estimates
from modern effect learners.

The resulting insights are informal and exploratory, reflecting a
division of labor in the clinical trials literature. In their
systematic review of subgroup methods,
\citet{ondra-2016-targeted-subgroups} separate \emph{confirmatory}
subgroup analysis, which considers a small number of prospectively
defined subgroups and protects the familywise error rate across them,
from \emph{exploratory} analysis, which ranges over many candidate
subgroups and where error rate control is typically not addressed at
all. Prior work has treated uplift models as exploratory, searching a
large space of subgroups and offering only suggestive rather than
definitive conclusions. Our procedure goes a step farther, drawing
rigorous conclusions from an exploratory search.

Although individual treatment effects are unobservable, random
assignment identifies average treatment effects in a fixed population.
The nodes of a decision tree are \emph{not} fixed populations, because
the selected nodes are a function of the data. The data-splitting
method of \citet{cox-1975-splitting} offers a solution: we divide the
dataset into a training and an evaluation set, fit models (uplift and
decision tree) on the training set, and draw conclusions from the
evaluation set. \citet{athey-2016-recursive-partitioning-hte} use this
strategy to derive unbiased estimates of treatment effects, a
procedure they call \emph{honest estimation}.

Their procedure emphasizes estimation rather than significance
testing, and does not offer simultaneous control of error rates.
Testing requires data splitting to make the subpopulations legitimate
targets of inference, but we also need the multiplicity adjustment of
\S\ref{sec:fwer} to make the resulting set of claims simultaneously
credible. We control the familywise error rate (FWER), the probability
of making at least one false statement, rather than the false
discovery rate (FDR), the fraction of statements made which are false.
While the FDR is valuable for exploratory work, it is a lower standard
of evidence than the FWER and is thus unsuitable for confirmatory
analysis: ``any desired conclusion can be almost surely inferred
without inflating the FDR'' \citep[page
  39]{dmitrienko-2009-multiple-testing}; see also
\citet{finner-2001-fdr}.

\section{Interweaving Hypotheses of Effect and Heterogeneity}
\label{sec:interweaving}

Each node of the decision tree, \(S\), denotes a subpopulation. Each
internal node has a split separating \(S\) into children \(L\) and
\(R.\) We consider two kinds of hypotheses.

\paragraph{Effect.} What can we infer about the average treatment effect in \(S\)?
We consider one-sided hypotheses:
\begin{displaymath}
  E^+_S : \tau_S \leq \delta_S
  \qquad\text{or}\qquad
  E^-_S : \tau_S \geq \delta_S,
\end{displaymath}
where \( \tau_S \) is the average treatment effect in \( S, \) and
\( \delta_S \) is a meaningful threshold we discuss below.
Rejecting \(E^+_S\) certifies---in the Frequentist sense of
``certify'' \citep{rubin-2004-teaching-stats}---the average treatment
effect in \( S \) is greater than \( \delta_S \). Rejecting \(E^-_S\)
certifies the other direction, which is often just as valuable to
know. \S\ref{sec:adaptive} describes how we select the direction.

Setting \( \delta_S = 0 \) allows us to identify the sign of the
average treatment effect in a group. A small negative \(\delta_i\)
implements a non-inferiority test, useful when ruling out catastrophic
harm is more important than proving a benefit. When applying the
treatment requires some resource, setting a suitable positive
\( \delta_i \) can determine whether the benefit justifies the
resource.

\paragraph{Heterogeneity.} Do the two subpopulations split at \( S\)
respond differently to treatment? Here also we consider one-sided
hypotheses:
\begin{displaymath}
  H^R_S : \tau_R \leq \tau_L
  \qquad\text{or}\qquad
  H^L_S : \tau_R \geq \tau_L .
\end{displaymath}
As with the effect hypotheses, \S\ref{sec:adaptive} discusses how to
decide which to test. Rejecting \(H^R_S\) certifies heterogeneity at
the split \emph{and} that the right child responds more positively.

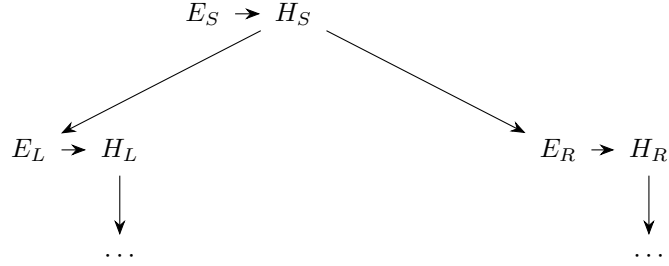
\begin{figure}[htbp]
\centering
\begin{tikzpicture}[
  every node/.style={font=\normalsize},
  arrow/.style={-{Stealth[length=2mm]}, shorten >=2pt, shorten <=2pt},
]
  \node (ES) at (-0.6,  0.0) {\(E_S\)};
  \node (HS) at ( 0.6,  0.0) {\(H_S\)};

  \node (EL) at (-2.9, -1.8) {\(E_L\)};
  \node (HL) at (-1.7, -1.8) {\(H_L\)};
  \node (ER) at ( 4.1, -1.8) {\(E_R\)};
  \node (HR) at ( 5.3, -1.8) {\(H_R\)};

  \node (CL) at (-1.7, -3.2) {\(\cdots\)};
  \node (CR) at ( 5.3, -3.2) {\(\cdots\)};

  \draw [arrow] (ES) -- (HS);
  \draw [arrow] (HS) -- (EL);
  \draw [arrow] (HS) -- (ER);
  \draw [arrow] (EL) -- (HL);
  \draw [arrow] (ER) -- (HR);
  \draw [arrow] (HL) -- (CL);
  \draw [arrow] (HR) -- (CR);
\end{tikzpicture}
\caption{\label{fig:interweaving}The interwoven hierarchy at one node
  and its two children. An effect hypothesis has one child: the
  heterogeneity hypothesis at the same node. A heterogeneity
  hypothesis has two: the effect hypotheses of the child nodes.}
\end{figure}

We interweave the two families into a single hierarchy to support
joint testing of both kinds of hypotheses. Each effect hypothesis has
one child, the heterogeneity hypothesis at the same node. Each
heterogeneity hypothesis has the effect hypotheses of the two child
nodes as its children, as depicted in Figure~\ref{fig:interweaving}.
Leaves carry an effect hypothesis and no heterogeneity hypothesis,
having no split beneath them.

\section{Adaptively Selecting Hypotheses}
\label{sec:adaptive}

Splitting the data does more than license inference after training. It
also lets us adapt the hypotheses to the data. We only learn something
when we reject a null hypothesis, so the goal is to reject as many
hypotheses as possible, and for the corresponding alternatives to be
informative.

We avoided two-sided hypotheses in the last section because the
alternative is not informative. Rejecting \( A: \theta = 0 \) only
proves \( \theta \neq 0 \) but does not determine the sign of
\( \theta \). Rejecting \( B: \theta \leq 0 \) \emph{is} informative,
certifying \( \theta > 0 \). A compelling strategy for two-sided
testing tests both one-sided possibilities at half the level (as a
Bonferroni adjustment), which \emph{does} identify on which side the
truth lies. But when we can anticipate the direction, we get more
power by skipping one of these tests and allocating the full level to
the one side.

In addition to providing the uplift model and the decision tree, the
training set supports just this, suggesting a hypothesis to test at
each node, \(E_S^+\) or \(E_S^-\), and \( H_S^R\) or \(H_S^L\). Write
\(\hat{\tau}_{S, \textrm{train}}\) for the average treatment effect
estimated in \( S \) on the training set. If
\(\hat{\tau}_{S,\textrm{train}} > \delta_S\) we select \(E_S^+\) to
test on the evaluation set; otherwise we select \(E_S^-.\) Similarly,
if \(\hat{\tau}_{R,\textrm{train}} > \hat{\tau}_{L,\textrm{train}} \),
we select \( H_S^R \) to test; otherwise, \( H_S^L \). Rejecting the
selected hypotheses certifies what we observed informally in the
training set.

Using one half of a sample to choose what to test and the other to
test it is not new. \citet{wasserman-2009-screen-and-clean} screen
variables on one half and clean up on the other, and
\citet{rubin-2006-sample-splitting} use a split to set cutoffs for a
multiple testing procedure. Both select \emph{which} hypotheses to
test. Here we apply the method to uplift modeling.

\section{Controlling the Familywise Error Rate}
\label{sec:fwer}

The methods proposed in this section operate on a finite or countable
collection of hypotheses, \(A_i, \, i \in \mathcal{I},\) arranged in a
tree-like hierarchy. The root node of this hierarchy will be called
\(A_1\). Each hypothesis \(i\) has zero or more children indexed by
\(\mathcal{J}_i.\) Other than \(A_1,\) all hypotheses have one parent
hypothesis. In our case, \(A_1\) corresponds to the average treatment
effect in the overall population.

Many existing techniques can take advantage of sequences or
hierarchies of hypotheses to increase statistical power against
certain alternatives. The Fixed Sequence Procedure
\citep{maurer-1995-fixed-sequence} and the Fallback Procedure
\citep{wiens-2003-fallback, wiens-2005-fallback-procedure} are two
well-known examples. The intuition behind these procedures is ``we pay
only for our failures; successes are free''. When we test and reject a
hypothesis, its level can be carried forward to later tests,
increasing their power, but when we fail to reject a hypothesis, its
level is ``spent''.

\citet{bretz-2009-graphical} and \citet{burman-2009-recycling}
developed graphical techniques generalizing the sequential procedures
to hierarchies of hypotheses, and
\citet{goeman-2010-sequential-rejection} later identified the general
principle. A graphical procedure is specified by an initial allocation
\(a\) of \(\alpha\) across the hypotheses, with
\(\sum_i a_i \leq \alpha,\) and a transition matrix \(G\) saying how a
rejected hypothesis's allocation is redistributed among the survivors.
In \S\ref{sec:trickle-down}, we discuss a special case of this
strategy relevant to the hierarchy we propose.

\subsection{Hierarchical Bonferroni Procedure}
\label{sec:bonferroni}

This section gives a Bonferroni adjustment controlling the
Familywise Error Rate, to measure the later methods against. Such an
adjustment assigns a weight to each hypothesis. Provided the weights
sum to no more than \(\alpha\), the adjustment controls the FWER at level
\(\alpha\). Let \(n_k\) be the number of hypotheses at depth \(k\)
of the hierarchy, with the root node occurring at depth 1. For
example, \(n_1 = 1,\) \(n_2 = | \mathcal{J}_1 |,\) where \(| \mathcal{J}_1 |\)
is the number of elements in \(\mathcal{J}_1\), and
\(n_3 = \sum_{j \in \mathcal{J}_1} | \mathcal{J}_j |.\)

The Hierarchical Bonferroni Procedure tests \(A_1\) at level
\(\alpha / 2.\) It tests the hypotheses at depth \(k\) at level
\(2^{-k} \cdot \alpha / n_k,\) \(k=2, \ldots.\) Since the levels
at each depth sum to \(2^{-k} \cdot \alpha,\) the sum of the
levels across the entire hierarchy is
\(\sum_k 2^{-k} \cdot \alpha \leq \alpha.\)

The Hierarchical Bonferroni Procedure thus controls the FWER at the
desired level, but the power corresponding to hypotheses lower in the
hierarchy decreases exponentially. As a special case, consider a
hierarchy where each hypothesis has two children. Then
\(n_k = 2^{k - 1},\) \(k=1, \ldots,\) and we test the hypotheses at
depth \(k\) at level \(2^{1-2 \cdot k} \cdot \alpha.\)

\subsection{Fixed Hierarchy Procedure}
\label{sec:fixed-hierarchy}

As with the Hierarchical Bonferroni Procedure, we have a collection of
hypotheses in a tree-like hierarchy. For this method, each hypothesis
\(A_i,\) other than \(A_1,\) has an associated weight,
\(w_i,\) with \(0 \leq w_i \leq 1\), and \(\sum_{j \in \mathcal{J}_i} w_j = 1\)
for all \(i.\)

The Fixed Hierarchy Procedure tests \(A_1\) at level \(\alpha_1 = \alpha.\)
If we fail to reject \(A_1\) (henceforth we say ``retain''), testing
stops; otherwise, we test each child at level \(w_j \cdot \alpha,\)
\(j \in \mathcal{J}_1.\) Progressing down the tree, if we test \(A_i\)
at level \(\alpha_i,\) and retain \(A_i,\) we do not test any
descendants of \(A_i\). Otherwise, we test each child of \(A_i\) at
level \( w_j \cdot \alpha_i, \) \(j \in \mathcal{J}_i.\)

\textbf{Remark:} the Fixed Sequence Procedure is a special case of the
Fixed Hierarchy Procedure, where each hypothesis has at most one
child.

This procedure is the hierarchical testing procedure of
\citet{meinshausen-2008-hierarchical}, with the sibling weights
generalized from proportional-to-size to arbitrary \( w_j. \) It is
the special case \(v_i \equiv 1\) of the procedure in
\S\ref{sec:trickle-down}. Strong control therefore follows from the
argument in \S\ref{sec:trickle-down}.

To compare the performance of the Fixed Hierarchy Procedure with
the Hierarchical Bonferroni Procedure, consider a hierarchy where
each hypothesis has two children, and \( w_i = 1/2 \) for all
\(i\). Assuming we reject all ancestors, the Fixed Hierarchy
Procedure tests a hypothesis at depth \(k\) at level
\(2^{1 - k} \cdot \alpha,\) compared to \(2^{1 - 2 \cdot k} \cdot \alpha\)
for the Hierarchical Bonferroni Procedure. The Fixed Hierarchy
Procedure can therefore be exponentially more powerful. It buys
power with flexibility: retain the first hypothesis and it rejects
\emph{nothing}, including hypotheses the Hierarchical Bonferroni
Procedure would have rejected. The method of the next section lets
testing continue past a retention, at lower power.

One property of both Hierarchical Bonferroni and Fixed Hierarchy is
that the level at which we test a hypothesis depends only on its
depth. We may grow the tree another level without changing the level
at which we test any existing hypothesis. Thus there is no reason to
control the tree depth when constructing the tree on the training set.
The power available to tests deep in the tree decreases exponentially,
so at some point, we will retain \emph{all} hypotheses below a certain
depth. We may prune these nodes, leaving a shallower tree emphasizing
the certified claims.

\subsection{Trickle Down Procedure}
\label{sec:trickle-down}

As before, each hypothesis has a weight, \(w_i.\) Each hypothesis now
also has a \emph{stake,} \(0 < v_i \leq 1.\) We mean the name in the
sense of a wager: something risked. We also need the \emph{available
level,} which the stake converts into the level at which we test a
hypothesis. If \(A_i\) has available level \(\alpha_i,\) we test
it at level \( v_i \cdot \alpha_i. \)

The Trickle Down Procedure assigns \(A_1\) to have available level
\(\alpha_1 := \alpha,\) testing it at level \(v_1 \cdot \alpha\). The
available level of hypotheses farther down in the hierarchy depends on
which hypotheses farther up in the hierarchy we rejected. Suppose
\(A_i\) has available level \(\alpha_i\) (and is thus tested at
level \(v_i \cdot \alpha_i\) ). We assign each child an available
level depending on whether we rejected the parent:
\begin{displaymath}
    \alpha_j := \begin{cases}
    w_j \cdot \alpha_i & A_i \textrm{ rejected} \\
    w_j \cdot (1 - v_i ) \cdot \alpha_i & A_i \textrm{ retained},
    \end{cases}
\end{displaymath}
\(j \in \mathcal{J}_i.\)

When we test \(A_i\), we wager only a portion, \(v_i,\) of the
available level, \(\alpha_i,\) testing \(A_i\) at level
\(v_i \cdot \alpha_i\). If we reject \(A_i,\) we split the entire
available level, \(\alpha_i,\) among the children. If we fail to
reject \(A_i,\) the stake is forfeit, but the unwagered portion,
\((1 - v_i) \cdot \alpha_i,\) is still available for continued
testing.

\textbf{Remark:} the Fixed Hierarchy Procedure is a special case of
the Trickle Down Procedure, where \(v_i = 1\) for all \(i.\) The Fixed
Hierarchy Procedure stakes the entire available level at each
opportunity. The Fallback Procedure is analogous to the Trickle Down
Procedure when each hypothesis has at most one child, but with
different weights.

This procedure is an instance of the graphical approach of
\citet{bretz-2009-graphical}, developed independently and in parallel
by \citet{burman-2009-recycling}, and strong control of the familywise
error rate follows from their results. The proof below is direct, and
short enough to be worth giving in the notation of this paper.

We first record that the procedure is monotone in the available
level.

\begin{lemma}
\label{lem:monotone}
Decreasing the available level at a node can only shrink the set of
hypotheses rejected at or below it.
\end{lemma}

\begin{proof}
Run the procedure from node \(i\) twice, with available levels
\(a \leq a'.\) The node is tested at \(v_i a\) and \(v_i a',\) so if
it is rejected under \(a\) it is rejected under \(a'.\) Each child
then receives \((1 - v_i) w_j a\) or \(w_j a\) in the first run,
according to whether \(A_i\) was retained or rejected, and the
corresponding quantity with \(a'\) in the second. In every case the
child's available level in the first run is no larger than in the
second, because \(a \leq a'\) and \(1 - v_i \leq 1.\) Descending the
subtree gives the claim.
\end{proof}

\begin{proposition}
\label{prop:fwer}
The probability the Trickle Down Procedure rejects at least one true
hypothesis is at most \(\alpha.\)
\end{proposition}

\begin{proof}
For a node \(i\) and a level \(a > 0,\) let \(F_i(a)\) be the event
that the procedure, run on the subtree rooted at \(i\) with available
level \(a,\) rejects at least one true hypothesis. Each such run is a
function of the same \(p\)-values, so these events are all defined on
one probability space. We show
\(\mathrm{Pr}\left\{ F_i(a) \right\} \leq a\) by induction up from the
leaves; the proposition is the case \(i = 1,\) \(a = \alpha.\)

If \(i\) is a leaf and \(A_i\) is false, \(F_i(a)\) is empty.
If \(A_i\) is true, \(F_i(a)\) is the event that the test at \(i\)
rejects, which has probability at most \(v_i a \leq a.\)

Now let \(i\) have children \(\mathcal{J}_i,\) and suppose the claim
holds for each child. Suppose first that \(A_i\) is true. Rejecting it
is itself a false rejection, and when it is retained each child
receives available level \((1 - v_i) w_j a,\) so
\begin{displaymath}
  F_i(a) \;\subseteq\; \left\{ A_i \textrm{ rejected} \right\}
  \;\cup\; \bigcup_{j \in \mathcal{J}_i}
  F_j\bigl( (1 - v_i) \, w_j \, a \bigr).
\end{displaymath}
The union bound and the inductive hypothesis give
\begin{displaymath}
  \mathrm{Pr}\left\{ F_i(a) \right\} \;\leq\; v_i \, a
  + \sum_{j \in \mathcal{J}_i} (1 - v_i) \, w_j \, a
  \;=\; v_i \, a + (1 - v_i) \, a \;=\; a .
\end{displaymath}

Suppose instead that \(A_i\) is false. Rejecting it is not an error,
and each child receives available level either \(w_j a\) or
\((1 - v_i) w_j a,\) both at most \(w_j a.\) By
Lemma~\ref{lem:monotone},
\begin{displaymath}
  F_i(a) \;\subseteq\; \bigcup_{j \in \mathcal{J}_i} F_j( w_j \, a ) ,
\end{displaymath}
so \(\mathrm{Pr}\left\{ F_i(a) \right\} \leq
\sum_{j \in \mathcal{J}_i} w_j \, a = a.\)
\end{proof}

The Trickle Down Procedure is more flexible than the Fixed Hierarchy
Procedure and can be more powerful than the Hierarchical Bonferroni
Procedure. To see this, suppose each hypothesis has two children, and
let \(w_i = v_i = 1/2\) for all \(i\). The level at which we test
hypotheses, other than the root hypothesis, depends on whether we
reject earlier hypotheses.

Consider then two possibilities: first that we have rejected all
hypotheses, second that we have retained all hypotheses.
Table~\ref{tab:three-methods} shows the level at which we test a
hypothesis at a particular depth. When we retain all hypotheses, the
Trickle Down Procedure is no better (and no worse) than the
Hierarchical Bonferroni Procedure. When we reject a hypothesis, we
test later hypotheses at higher levels than with Hierarchical
Bonferroni.

\begin{table}[htbp]
\caption{\label{tab:three-methods}Level at which a hypothesis is
  tested, for various depths, by the three methods.}
\centering
\begin{tabular}{rrrrr}
Depth & Hier.\ Bonferroni & Trickle Down & Trickle Down & Fixed Hierarchy\\
 &  & (all retained) & (all rejected) & \\
\hline
1 & \(\alpha/2\) & \(\alpha/2\) & \(\alpha/2\) & \(\alpha\)\\
2 & \(\alpha/8\) & \(\alpha/8\) & \(\alpha/4\) & \(\alpha/2\)\\
3 & \(\alpha/32\) & \(\alpha/32\) & \(\alpha/8\) & \(\alpha/4\)\\
4 & \(\alpha/128\) & \(\alpha/128\) & \(\alpha/16\) & \(\alpha/8\)\\
\end{tabular}
\end{table}

\begin{remark}
  One consequence of the tree parameterization is that it only
  exhausts \(\alpha\) when all leaves have \( v = 1\). Since Fixed
  Hierarchy sets \( v = 1 \) for all nodes, it is \( \alpha \)-exhaustive.
\end{remark}

\section{Confidence Bounds for Retained Hypotheses}
\label{sec:bounds}

As we get deeper in the tree, available level decreases and we may
retain some hypotheses. This section describes a consolation prize: a
confidence bound corresponding to a weakened form of the hypothesis.

Let \( \theta \geq 0 \). Let the \emph{weakened} version of a hypothesis
be as follows:
\begin{itemize}
   \item{\( E_S^+(\theta): \tau_S \leq \delta_S - \theta \)}
   \item{\( E_S^-(\theta): \tau_S \geq \delta_S + \theta \)}
   \item{\( H_S^R(\theta): \tau_R \leq \tau_L - \theta \)}
   \item{\( H_S^L(\theta): \tau_R \geq \tau_L + \theta \)}
\end{itemize}
Weaker versions (having larger values of \( \theta \)) are easier to
reject. At \( \theta = 0, \) the weakened version corresponds to the
non-weakened version.

We may fail to reject a (non-weakened) hypothesis, but we can report the
smallest \( \theta \) at which we can reject the weakened hypothesis.
For example, when testing \( E_S^+ \) with \( \delta_S = 0 \), we may
not be able to prove the treatment is beneficial, but we can calculate
a bound on how harmful it is.

\begin{proposition}
\label{prop:bounds}
Run the Trickle Down Procedure on the hierarchy of
\S\ref{sec:interweaving}, and let each node report the smallest
\( \theta \geq 0 \) where the weakened hypothesis is rejected. The
probability that any node falsely rejects a weakened hypothesis is
still at most \(\alpha.\)
\end{proposition}

\begin{proof}
Fix a node and write \(A(\theta)\) for the weakened hypotheses it
tests. \(A(0)\) is the hypothesis the Trickle Down Procedure tests.
Let \(\theta_0\) be the margin by which \(A(0)\) holds: for example,
\(\delta_S - \tau_S\) for \(E^+_S(\theta).\) In each case
\(A(\theta)\) is true if and only if \(\theta \leq \theta_0\). \( A(\theta_0) \)
is true, holding with equality. When \( \theta_0 \geq 0 \), \( A(0) \)
is true. Because we consider only \( \theta \geq 0 \), when \(
\theta_0 < 0 \), the node never considers a true hypothesis and can
reject no true hypothesis, so assume \( \theta_0 \geq 0 \).

The \(p\)-value decreases in \(\theta\), so the node rejects for
\( \theta \geq \theta^\star \). If the node rejects \( A(\theta_0) \)
it has rejected a true hypothesis. Conversely, if the node rejects a
true hypothesis, then \( A(\theta) \) is true for some
\( \theta \geq \theta^\star \); any such \( \theta \) is at most
\( \theta_0 \), so \( \theta^\star \leq \theta_0 \) and the node
rejects \( A(\theta_0) \). The event ``the node rejects a true
hypothesis'' is therefore the event ``the node rejects
\( A(\theta_0) \)''.

We test all hypotheses \( A(\theta) \) for \( \theta \geq 0 \),
including \( A(\theta_0), \) at the same level at which we test
\( A(0) \), so the probability of falsely rejecting \( A(\theta_0) \)
and the probability of falsely rejecting \( A(0) \) are both at most
that level.

The enlarged family therefore presents the procedure of
\S\ref{sec:trickle-down} with the same hierarchy, the same weights and
stakes, and one valid test at each node, which is all
Proposition~\ref{prop:fwer} requires. Only the statement each node
makes has changed.
\end{proof}

Less formally: every way a node can be wrong about its bound is the
same failure, rejecting \( A(\theta_0) \), and our procedure controls
the probability of a false rejection. Reporting the bound adds no new
opportunity to be wrong. The restriction \(\theta \geq 0 \) is
necessary. We cannot compute a confidence bound corresponding to a
rejected hypothesis without finding a threshold at which we retain the
\emph{strengthened} hypothesis, which prevents trickling its power
down to later hypotheses. The asymmetry restates the principle:
successes are free; failures are not.

\section{Simulation Study}
\label{sec:simulation}

We investigated two aspects of the procedures: whether they control
the familywise error rate (\S\ref{sec:sim-fwer}), and their ability to
certify true effects (\S\ref{sec:sim-adaptive}). We begin with a
single worked run.

\subsection{A worked example}
\label{sec:sim-example}

We generated 10,000 individuals, each described by ten features: a
continuous outcome predictor \(x_0,\) a three-level categorical
outcome predictor \(x_1,\) two binary effect moderators \(x_2\) and
\(x_3,\) and six independent standard-normal noise variables
\(x_4, \ldots, x_9\). The baseline outcome was
\(Y(0) = x_0 + 0.3 x_1 + \varepsilon_0\) with
\(\varepsilon_0 \sim \mathcal{N}(0, 9),\) and
\(Y(1) = Y(0) + \tau(x) + \varepsilon_1\) with \(\varepsilon_1\)
independent of \(\varepsilon_0.\) The true treatment effect depended
only on the two binary moderators:

\begin{table}[h]
\centering
\begin{tabular}{rrr}
\(x_2\) & \(x_3\) & \(\tau\)\\
\hline
0 & 0 & \(-\)1.00\\
0 & 1 & 1.20\\
1 & 0 & 2.00\\
1 & 1 & 1.50\\
\end{tabular}
\end{table}

We split the population evenly across the four moderator levels, so
the population average treatment effect was \(+0.925\). The average is
misleading: the \( x_2 = 0, \) \( x_3 = 0 \) subpopulation experienced
a negative treatment effect masked by the others. The average
treatment effect among people having \(x_2 = 0\) is \( +0.1,\) but the
average conceals a strong positive effect just canceling out a strong
negative effect. We assigned each individual to treatment or control
with equal probability, recorded the corresponding potential outcome,
and split the sample into training and evaluation halves.

On the training half we fit a T-learner: separate regression models
\(\hat{\mu}_0\) and \(\hat{\mu}_1\) on the two arms. We estimated
individual treatment effects in the training set as
\(\hat{\mu}_1(x) - y_{\text{obs}}\) for controls and
\(y_{\text{obs}} - \hat{\mu}_0(x)\) for the treated. We fit a
depth-three regression tree to those individual effects, again in the
training set, using all ten features, with leaves constrained to hold
at least a hundredth of the sample. The depth constraint is just for
legibility in this example; the method works with arbitrarily deep
trees and offers a disciplined pruning strategy. The leaf constraint
ensures the sample size is large enough to run the associated tests.

We applied the interwoven hierarchy of \S\ref{sec:interweaving} to
the resulting tree, with \(\alpha = 0.05,\) stake \(v_i = 0.5\) at
every node, equal weights between siblings, and
threshold \(\delta = 0\) throughout. Directions were chosen on the
training half, as in \S\ref{sec:adaptive}. All tests used the
evaluation half. Each node's hypothesis is tested by a one-sided
Welch \(t\) test on the relevant contrast---the difference of arm
means for an effect hypothesis, and the difference-in-differences
\((\bar{y}^R_T - \bar{y}^R_C) - (\bar{y}^L_T - \bar{y}^L_C)\) for a
heterogeneity one---with unpooled variances and Welch--Satterthwaite
degrees of freedom. The bounds of \S\ref{sec:bounds} come from
inverting those same tests. Any test of the same contrasts at the
desired level would serve.

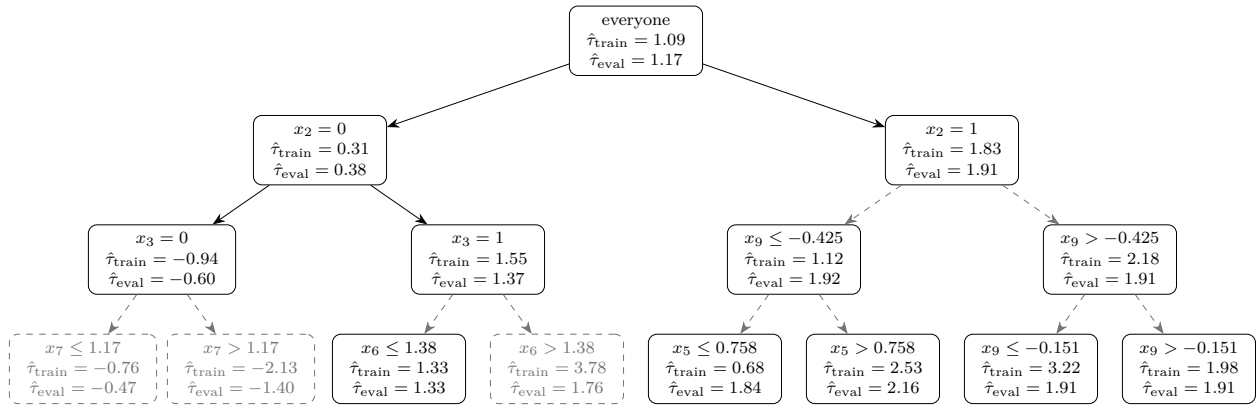
\begin{figure}[htbp]
\centering
\resizebox{\linewidth}{!}{
\begin{tikzpicture}[
  every node/.style={draw, rounded corners, inner sep=2pt, font=\footnotesize},
  established/.style={solid, draw=black, text=black},
  unestablished/.style={dashed, text=black!55, draw=black!55},
  unestablished edge/.style={dashed, draw=black!55},
  level 1/.style={sibling distance=104mm, level distance=18mm},
  level 2/.style={sibling distance=52mm, level distance=18mm},
  level 3/.style={sibling distance=26mm, level distance=18mm},
  edge from parent/.style={draw, -{Stealth[length=2mm]}},
]
\node[established] {\begin{tabular}{c}everyone\\$\hat{\tau}_{\mathrm{train}} = 1.09$\\$\hat{\tau}_{\mathrm{eval}} = 1.17$\end{tabular}}
  child {
    node[established] {\begin{tabular}{c}$x_{2} = 0$\\$\hat{\tau}_{\mathrm{train}} = 0.31$\\$\hat{\tau}_{\mathrm{eval}} = 0.38$\end{tabular}}
      child {
        node[established] {\begin{tabular}{c}$x_{3} = 0$\\$\hat{\tau}_{\mathrm{train}} = {-}0.94$\\$\hat{\tau}_{\mathrm{eval}} = {-}0.60$\end{tabular}}
          child {
            node[unestablished] {\begin{tabular}{c}$x_{7} \leq 1.17$\\$\hat{\tau}_{\mathrm{train}} = {-}0.76$\\$\hat{\tau}_{\mathrm{eval}} = {-}0.47$\end{tabular}}
            edge from parent[unestablished edge]
          }
          child {
            node[unestablished] {\begin{tabular}{c}$x_{7} > 1.17$\\$\hat{\tau}_{\mathrm{train}} = {-}2.13$\\$\hat{\tau}_{\mathrm{eval}} = {-}1.40$\end{tabular}}
            edge from parent[unestablished edge]
          }
      }
      child {
        node[established] {\begin{tabular}{c}$x_{3} = 1$\\$\hat{\tau}_{\mathrm{train}} = 1.55$\\$\hat{\tau}_{\mathrm{eval}} = 1.37$\end{tabular}}
          child {
            node[established] {\begin{tabular}{c}$x_{6} \leq 1.38$\\$\hat{\tau}_{\mathrm{train}} = 1.33$\\$\hat{\tau}_{\mathrm{eval}} = 1.33$\end{tabular}}
            edge from parent[unestablished edge]
          }
          child {
            node[unestablished] {\begin{tabular}{c}$x_{6} > 1.38$\\$\hat{\tau}_{\mathrm{train}} = 3.78$\\$\hat{\tau}_{\mathrm{eval}} = 1.76$\end{tabular}}
            edge from parent[unestablished edge]
          }
      }
  }
  child {
    node[established] {\begin{tabular}{c}$x_{2} = 1$\\$\hat{\tau}_{\mathrm{train}} = 1.83$\\$\hat{\tau}_{\mathrm{eval}} = 1.91$\end{tabular}}
      child {
        node[established] {\begin{tabular}{c}$x_{9} \leq -0.425$\\$\hat{\tau}_{\mathrm{train}} = 1.12$\\$\hat{\tau}_{\mathrm{eval}} = 1.92$\end{tabular}}
          child {
            node[established] {\begin{tabular}{c}$x_{5} \leq 0.758$\\$\hat{\tau}_{\mathrm{train}} = 0.68$\\$\hat{\tau}_{\mathrm{eval}} = 1.84$\end{tabular}}
            edge from parent[unestablished edge]
          }
          child {
            node[established] {\begin{tabular}{c}$x_{5} > 0.758$\\$\hat{\tau}_{\mathrm{train}} = 2.53$\\$\hat{\tau}_{\mathrm{eval}} = 2.16$\end{tabular}}
            edge from parent[unestablished edge]
          }
        edge from parent[unestablished edge]
      }
      child {
        node[established] {\begin{tabular}{c}$x_{9} > -0.425$\\$\hat{\tau}_{\mathrm{train}} = 2.18$\\$\hat{\tau}_{\mathrm{eval}} = 1.91$\end{tabular}}
          child {
            node[established] {\begin{tabular}{c}$x_{9} \leq -0.151$\\$\hat{\tau}_{\mathrm{train}} = 3.22$\\$\hat{\tau}_{\mathrm{eval}} = 1.91$\end{tabular}}
            edge from parent[unestablished edge]
          }
          child {
            node[established] {\begin{tabular}{c}$x_{9} > -0.151$\\$\hat{\tau}_{\mathrm{train}} = 1.98$\\$\hat{\tau}_{\mathrm{eval}} = 1.91$\end{tabular}}
            edge from parent[unestablished edge]
          }
        edge from parent[unestablished edge]
      }
  }
;
\end{tikzpicture}}
\caption{\label{fig:fitted-tree}The fitted tree, with the effect
  estimated in each node on both halves: \(\hat{\tau}_{\mathrm{train}}\)
  chose the direction of every test, and
  \(\hat{\tau}_{\mathrm{eval}}\) is the estimate the test was run on.}
\end{figure}

Figure~\ref{fig:fitted-tree} shows the tree as fit on the training
set. It correctly identified \(x_2\) as the most important predictor
of effect heterogeneity, but detected \( x_3 \) as an effect moderator
only on the \(x_2=0\) branch. On the \(x_2=1\) branch, we found
spurious heterogeneity with respect to \( x_9. \)

Nodes and arrows can be solid or dashed, indicating whether we
rejected the corresponding hypothesis with the specified FWER. From
the top: the overall average treatment effect was positive (s.s.); the
treatment effect among \(x_2=1\) was higher than for \(x_2=0\) (s.s.);
among people having \(x_2=0,\) the treatment effect was higher for
people having \(x_3=1\) than for \(x_3=0\) (s.s.); the average
treatment effect among people having \(x_2=0, \, x_3=0\) was
\textbf{negative} (s.s.), but positive for \(x_2=0, \, x_3=1\) (s.s.),
and so forth. The spurious heterogeneity with respect to \( x_9 \) was
not statistically significant. There are no Type-I errors shown in
Figure~\ref{fig:fitted-tree}, but we failed to identify heterogeneity
with respect to \( x_3 \) among the \( x_2=1 \) subpopulation.

Figure~\ref{fig:levels} shows the level at which each hypothesis is
tested, the p-value, and for retained hypotheses, the ``consolation
prize'' confidence bound. The figure shows, for example, that even
though we cannot prove the treatment is beneficial for the
\( x_2 = 0, \) \( x_3 = 1, \) \( x_6 > 1.38 \) group, we can bound how
harmful it might be.

\begin{figure}[htbp]
\centering
\resizebox{\linewidth}{!}{
\begin{tikzpicture}[
  hyp/.style={draw, rounded corners, inner sep=3pt, font=\footnotesize},
  erej/.style={hyp, solid, draw=metablue, text=metablue},
  eret/.style={hyp, dashed, draw=metablue, text=metablue},
  hrej/.style={hyp, solid, draw=metaorange, text=metaorange},
  hret/.style={hyp, dashed, draw=metaorange, text=metaorange},
  plain/.style={draw=none, font=\scriptsize},
  flow/.style={-{Stealth[length=2mm]}, shorten >=2pt, shorten <=2pt},
  earrow/.style={flow, draw=metablue},
  edasharrow/.style={flow, dashed, draw=metablue},
  harrow/.style={flow, draw=metaorange},
  hdasharrow/.style={flow, dashed, draw=metaorange},
]
\node[erej] (n14e) at (9.30,-0.00) {\begin{tabular}{c}$0.0250$\\$p<0.001$\end{tabular}};
\node[hrej] (n14h) at (12.40,-0.00) {\begin{tabular}{c}$0.0250$\\$p<0.001$\end{tabular}};
\node[erej] (n6e) at (3.10,-3.00) {\begin{tabular}{c}$0.0125$\\$p=0.007$\end{tabular}};
\node[hrej] (n6h) at (6.20,-3.00) {\begin{tabular}{c}$0.0125$\\$p<0.001$\end{tabular}};
\node[erej] (n13e) at (15.50,-3.00) {\begin{tabular}{c}$0.0125$\\$p<0.001$\end{tabular}};
\node[hret] (n13h) at (18.60,-3.00) {\begin{tabular}{c}$0.0125$\\$p=0.512$\\$\tau_R - \tau_L \geq -0.727$\end{tabular}};
\node[erej] (n2e) at (0.00,-6.00) {\begin{tabular}{c}$0.0063$\\$p=0.003$\end{tabular}};
\node[hret] (n2h) at (3.10,-6.00) {\begin{tabular}{c}$0.0063$\\$p=0.084$\\$\tau_R - \tau_L \leq 0.770$\end{tabular}};
\node[erej] (n5e) at (6.20,-6.00) {\begin{tabular}{c}$0.0063$\\$p<0.001$\end{tabular}};
\node[hret] (n5h) at (9.30,-6.00) {\begin{tabular}{c}$0.0063$\\$p=0.291$\\$\tau_R - \tau_L \geq -1.539$\end{tabular}};
\node[erej] (n9e) at (12.40,-6.00) {\begin{tabular}{c}$0.0031$\\$p<0.001$\end{tabular}};
\node[hret] (n9h) at (15.50,-6.00) {\begin{tabular}{c}$0.0031$\\$p=0.307$\\$\tau_R - \tau_L \geq -1.391$\end{tabular}};
\node[erej] (n12e) at (18.60,-6.00) {\begin{tabular}{c}$0.0031$\\$p<0.001$\end{tabular}};
\node[hret] (n12h) at (21.70,-6.00) {\begin{tabular}{c}$0.0031$\\$p=0.496$\\$\tau_R - \tau_L \leq 1.454$\end{tabular}};
\node[eret] (n0e) at (0.00,-9.00) {\begin{tabular}{c}$0.0031$\\$p=0.024$\\$\tau \leq 0.176$\end{tabular}};
\node[eret] (n1e) at (3.10,-9.00) {\begin{tabular}{c}$0.0031$\\$p=0.014$\\$\tau \leq 0.359$\end{tabular}};
\node[erej] (n3e) at (6.20,-9.00) {\begin{tabular}{c}$0.0031$\\$p<0.001$\end{tabular}};
\node[eret] (n4e) at (9.30,-9.00) {\begin{tabular}{c}$0.0031$\\$p=0.010$\\$\tau \geq -0.317$\end{tabular}};
\node[erej] (n7e) at (12.40,-9.00) {\begin{tabular}{c}$0.0016$\\$p<0.001$\end{tabular}};
\node[erej] (n8e) at (15.50,-9.00) {\begin{tabular}{c}$0.0016$\\$p<0.001$\end{tabular}};
\node[erej] (n10e) at (18.60,-9.00) {\begin{tabular}{c}$0.0016$\\$p<0.001$\end{tabular}};
\node[erej] (n11e) at (21.70,-9.00) {\begin{tabular}{c}$0.0016$\\$p<0.001$\end{tabular}};
\node[plain, anchor=south] at ([xshift=0.00cm, yshift=1.5mm] n14e.north) {everyone};
\node[plain, anchor=south] at ([xshift=-0.50cm, yshift=1.5mm] n6e.north) {$x_{2} = 0$};
\node[plain, anchor=south] at ([xshift=0.50cm, yshift=1.5mm] n13e.north) {$x_{2} = 1$};
\node[plain, anchor=south] at ([xshift=-0.50cm, yshift=1.5mm] n2e.north) {$x_{3} = 0$};
\node[plain, anchor=south] at ([xshift=0.50cm, yshift=1.5mm] n5e.north) {$x_{3} = 1$};
\node[plain, anchor=south] at ([xshift=-0.50cm, yshift=1.5mm] n9e.north) {$x_{9} \leq -0.425$};
\node[plain, anchor=south] at ([xshift=0.50cm, yshift=1.5mm] n12e.north) {$x_{9} > -0.425$};
\node[plain, anchor=south] at ([xshift=-0.50cm, yshift=1.5mm] n0e.north) {$x_{7} \leq 1.17$};
\node[plain, anchor=south] at ([xshift=0.50cm, yshift=1.5mm] n1e.north) {$x_{7} > 1.17$};
\node[plain, anchor=south] at ([xshift=-0.50cm, yshift=1.5mm] n3e.north) {$x_{6} \leq 1.38$};
\node[plain, anchor=south] at ([xshift=0.50cm, yshift=1.5mm] n4e.north) {$x_{6} > 1.38$};
\node[plain, anchor=south] at ([xshift=-0.50cm, yshift=1.5mm] n7e.north) {$x_{5} \leq 0.758$};
\node[plain, anchor=south] at ([xshift=0.50cm, yshift=1.5mm] n8e.north) {$x_{5} > 0.758$};
\node[plain, anchor=south] at ([xshift=-0.50cm, yshift=1.5mm] n10e.north) {$x_{9} \leq -0.151$};
\node[plain, anchor=south] at ([xshift=0.50cm, yshift=1.5mm] n11e.north) {$x_{9} > -0.151$};
\draw[earrow] (n14e) -- (n14h);
\draw[harrow] (n14h.south) -- (n6e.north east);
\draw[harrow] (n14h.south) -- (n13e.north west);
\draw[earrow] (n6e) -- (n6h);
\draw[harrow] (n6h.south) -- (n2e.north east);
\draw[harrow] (n6h.south) -- (n5e.north west);
\draw[earrow] (n13e) -- (n13h);
\draw[hdasharrow] (n13h.south) -- (n9e.north east);
\draw[hdasharrow] (n13h.south) -- (n12e.north west);
\draw[earrow] (n2e) -- (n2h);
\draw[hdasharrow] (n2h.south) -- (n0e.north east);
\draw[hdasharrow] (n2h.south) -- (n1e.north west);
\draw[earrow] (n5e) -- (n5h);
\draw[hdasharrow] (n5h.south) -- (n3e.north east);
\draw[hdasharrow] (n5h.south) -- (n4e.north west);
\draw[earrow] (n9e) -- (n9h);
\draw[hdasharrow] (n9h.south) -- (n7e.north east);
\draw[hdasharrow] (n9h.south) -- (n8e.north west);
\draw[earrow] (n12e) -- (n12h);
\draw[hdasharrow] (n12h.south) -- (n10e.north east);
\draw[hdasharrow] (n12h.south) -- (n11e.north west);
\end{tikzpicture}}
\caption{\label{fig:levels}The interwoven hierarchy over the fitted
  tree, with the level each hypothesis was tested at, its
  \(p\)-value, and, where it was retained, the bound of
  \S\ref{sec:bounds}. Blue is an effect hypothesis and its bound is
  on \(\tau;\) orange is a heterogeneity hypothesis and its bound is
  on the contrast \(\tau_R - \tau_L.\) Solid is rejected, dashed
  retained. The \(p\)-values and bounds are from the held-out half.}
\end{figure}

These insights go well beyond reporting overall average treatment
effects. The casual observer may have drawn these same insights from a
decision tree, but without any statistical guarantees. The testing
procedure of \S\ref{sec:fwer} separates the reliable from the
spurious.

\subsection{Familywise error rate}
\label{sec:sim-fwer}

The remaining studies compare five procedures on the same interwoven
hierarchy: plain Bonferroni, which tests every hypothesis in the
family at \(\alpha/m\); hierarchical Bonferroni, which halves the
level at each depth; the fixed hierarchy procedure (\(v \equiv 1\));
the trickle down procedure with stake \(v \equiv 1/2\); and a
non-adaptive variant of the last, also at \(v \equiv 1/2,\)
forgoing direction selection entirely (every hypothesis is two-sided).
Only the last procedure ignores the selected directions; the other
four use them. Throughout, \(\alpha = 0.05\) and \(n = 4000.\)

Table~\ref{tab:sim-fwer} shows the simulated familywise error rate for
each procedure, the number of simulations where at least one true
hypothesis was rejected. Under the global null, every procedure
controls the familywise error rate, on a fixed partition and through a
fitted tree alike. The worst cell is the fixed hierarchy procedure at
\(0.0508\) (s.e.\ \(0.0035\)). Plain Bonferroni is the most
conservative, markedly so through a fitted tree.

Trickle down and hierarchical Bonferroni agree exactly, an artifact of
the definition of familywise error rate. At \(v \equiv 1/2\) with
equal sibling weights the trickle down allocation \emph{is} the
halve-at-each-depth allocation, so recycling is the only thing
separating the two procedures. Rejection triggers recycling, and under
the global null a rejection is already a familywise error, so no
amount of it can change whether one occurred.

The fitted-tree column shows no FWER inflation despite using an
estimated partition. We measure both columns on the same populations,
so the comparison between them is paired. Every paired difference is
within noise except plain Bonferroni's, which falls by \(0.0103\)
(s.e.\ \(0.0042\)) through the fitted tree. The direction is
conservative, so the control claim holds.

\begin{table}[htbp]
\caption{\label{tab:sim-fwer}Familywise error rate under the global
  null, \(\alpha = 0.05\), with its binomial standard error. Left: a
  partition fixed in advance, 4000 replications. Right: a depth-three
  tree distilled from a T-learner on the training half, the same depth
  as the worked example. Both columns are measured on the same 4000
  populations, so the difference between them is paired.}
\centering
\begin{tabular}{lrr}
Procedure & Fixed partition & Fitted tree\\
\hline
Bonferroni & $0.0447$ (s.e.\ $0.0033$) & $0.0345$ (s.e.\ $0.0029$)\\
Hierarchical Bonferroni & $0.0447$ (s.e.\ $0.0033$) & $0.0457$ (s.e.\ $0.0033$)\\
Fixed hierarchy & $0.0508$ (s.e.\ $0.0035$) & $0.0508$ (s.e.\ $0.0035$)\\
Trickle down & $0.0447$ (s.e.\ $0.0033$) & $0.0457$ (s.e.\ $0.0033$)\\
Non-adaptive & $0.0460$ (s.e.\ $0.0033$) & $0.0435$ (s.e.\ $0.0032$)\\
\end{tabular}

\end{table}

\subsection{Statistical power}
\label{sec:sim-adaptive}

The last section showed all procedures control the FWER at the
specified level. This section investigates statistical power: how many
true claims we certify, as a function of effect size. We scale two
effect configurations by a factor to show how the procedure works on
small and large effect sizes.

The first configuration is the running example of
\S\ref{sec:sim-example}: the treatment helps on average and helps
three quarters of the population, but harms the remaining quarter, so
the average is misleading. The second is the same configuration with
that quarter helped as well, so every effect points the same way.
Table~\ref{tab:sim-power} displays the results.

\begin{table}[htbp]
\caption{\label{tab:sim-power}Mean number of true claims certified,
  by effect scale, 4000 replications, with the standard error of each
  mean in parentheses. Top: the running example of
  \S\ref{sec:sim-example}, in which the treatment helps on average
  and harms one quarter. Bottom: the same configuration with that
  quarter helped instead. Ten true claims are available in each.}
\centering
\begin{tabular}{lrrrrrr}
Procedure & 0.10 & 0.15 & 0.20 & 0.30 & 0.45 & 1.00\\
\hline
\multicolumn{7}{l}{\emph{The running example: one quarter harmed (of 10)}}\\
Bonferroni & 0.13 (0.01) & 0.24 (0.01) & 0.43 (0.01) & 1.11 (0.02) & 2.86 (0.03) & 7.45 (0.01)\\
Hierarchical Bonferroni & 0.12 (0.01) & 0.23 (0.01) & 0.39 (0.01) & 0.96 (0.02) & 2.36 (0.02) & 6.68 (0.01)\\
Fixed hierarchy & 0.11 (0.01) & 0.20 (0.01) & 0.35 (0.01) & 0.84 (0.02) & 1.90 (0.02) & 3.60 (0.02)\\
Trickle down & 0.15 (0.01) & 0.28 (0.01) & 0.49 (0.02) & 1.25 (0.02) & 3.15 (0.03) & 7.58 (0.01)\\
Non-adaptive & 0.12 (0.01) & 0.22 (0.01) & 0.37 (0.01) & 0.93 (0.02) & 2.53 (0.03) & 7.36 (0.01)\\
\hline
\multicolumn{7}{l}{\emph{Every subgroup helped (of 10)}}\\
Bonferroni & 0.13 (0.01) & 0.24 (0.01) & 0.44 (0.01) & 1.13 (0.02) & 2.69 (0.02) & 6.42 (0.01)\\
Hierarchical Bonferroni & 0.13 (0.01) & 0.26 (0.01) & 0.46 (0.01) & 1.06 (0.02) & 2.19 (0.02) & 5.91 (0.02)\\
Fixed hierarchy & 0.15 (0.01) & 0.29 (0.01) & 0.48 (0.01) & 0.96 (0.01) & 1.65 (0.02) & 4.65 (0.03)\\
Trickle down & 0.16 (0.01) & 0.32 (0.01) & 0.57 (0.02) & 1.34 (0.02) & 2.86 (0.02) & 6.84 (0.02)\\
Non-adaptive & 0.13 (0.01) & 0.25 (0.01) & 0.42 (0.01) & 1.05 (0.02) & 2.42 (0.02) & 6.45 (0.02)\\
\end{tabular}

\end{table}

The non-adaptive variant forgoes direction selection in both families:
every hypothesis of both kinds is two-sided. Without a defensible
strategy for selecting the direction of a one-sided test, the
two-sided test is reasonable, but an increase in power is possible by
using the training set to select the direction. At small effect sizes,
where the adaptive strategy is more likely to choose the wrong side,
it still certifies more true effects than the two-sided variant.
Moreover, rejecting a two-sided hypothesis does not indicate on which
side the truth lies and can lead to Type III errors.
Table~\ref{tab:sim-power} credits both procedures for reaching a node,
so it understates the difference.

Plain Bonferroni does surprisingly well, but this reveals a limitation
of this evaluation strategy. Bonferroni gives less weight than the
other methods to hypotheses early in the hierarchy, and more weight to
the leaf hypotheses. There is only one root hypothesis, but many leaf
hypotheses, so testing the latter at higher levels leads to more
hypotheses rejected. This is an artifact of our shallow trees. In real
examples, we may grow deeper trees with no risk of sacrificing
reliability or power (except at the leaves) when using the
multiplicity adjustments of \S\ref{sec:fwer}. Plain Bonferroni is
useless in deep trees whose hypothesis count doubles with each new
level.

In our setting, hypotheses earlier in the hierarchy are more
important; rejecting a leaf hypothesis is less valuable than rejecting
the root hypothesis, but this is not reflected in
Table~\ref{tab:sim-power}. Table~\ref{tab:sim-depth} shows how the
procedures perform at each level. Fixed hierarchy has the highest
power at the root, but plummets thereafter (as expected). Trickle down
is almost as powerful at the root, but degrades more gracefully as we
go deeper in the tree. Its advantage over Hierarchical Bonferroni
mirrors Table~\ref{tab:three-methods}.

\begin{table}[htbp]
\caption{\label{tab:sim-depth}The same sweep at effect scale
  \(0.45\), stratified by the depth of the claim. Depth zero is the
  root. The totals of Table~\ref{tab:sim-power} are these rows summed,
  which weights each depth by how many nodes it holds.}
\centering
\begin{tabular}{lrrr}
Procedure & Root & Depth 1 & Depth 2\\
\hline
\multicolumn{4}{l}{\emph{The running example: one quarter harmed}}\\
\quad\emph{claims available} & (of 2) & (of 4) & (of 4)\\
Bonferroni & 0.77 & 1.06 & 1.02\\
Hierarchical Bonferroni & 1.13 & 0.88 & 0.36\\
Fixed hierarchy & 1.30 & 0.56 & 0.03\\
Trickle down & 1.20 & 1.04 & 0.90\\
Non-adaptive & 0.99 & 0.86 & 0.68\\
\hline
\multicolumn{4}{l}{\emph{Every subgroup helped}}\\
\quad\emph{claims available} & (of 2) & (of 4) & (of 4)\\
Bonferroni & 0.86 & 0.90 & 0.93\\
Hierarchical Bonferroni & 1.07 & 0.78 & 0.34\\
Fixed hierarchy & 1.28 & 0.36 & 0.02\\
Trickle down & 1.14 & 0.94 & 0.79\\
Non-adaptive & 1.02 & 0.79 & 0.61\\
\end{tabular}

\end{table}

\section{Summary}
\label{sec:summary}

Uplift models are often used to target a treatment where it will have
the most benefit. The techniques of this paper can help interpret and
validate such a strategy. Average treatment effects do not always tell
the whole story and may mask a negative effect for some people. A
responsible analyst will explore this possibility before administering
the treatment to everyone.

Decision trees lead to interpretable descriptions of how the treatment
effects vary across the population. Data splitting and multiplicity
adjustments lead to honest estimates of treatment effects and
trustworthy certification.

Table~\ref{tab:sim-depth} showed Fixed Hierarchy had the best power
for testing the root hypothesis, while plain Bonferroni had the best
power for testing the leaf hypotheses. Analysts may use the Trickle
Down stake, \( v, \) to tune the power available throughout the tree.
A larger stake provides more power early in the hierarchy, while a
smaller stake saves more power for later. The training data provide
treatment effect estimates in each cell. It may be possible to use
these estimates as guidance on how much power we need for each
hypothesis. We could use this analysis to tune the stake for each
hypothesis. We have not yet explored this.

Estimates and conclusions depend on the randomized split into training
and evaluation sets. This may lead two analysts to draw different
conclusions about treatment effects. We have not explored the
stability of these conclusions under repeated splits; this variability
is not captured in the significance levels or confidence bounds.

The hierarchy implies logical relations the procedure ignores. The
average treatment effect in a node is a weighted average of the
effects in the children, so a hypothesis about the parent can be false
only if corresponding hypotheses about at least one child are false.
Shaffer's method \citep{shaffer-1986-modified} exploits such
relations, but is not obviously applicable here because the directions
of the tests in parent and child may differ.

\bibliographystyle{plainnat}
\bibliography{citations}

\begin{thebibliography}{25}
\providecommand{\natexlab}[1]{#1}
\providecommand{\url}[1]{\texttt{#1}}
\expandafter\ifx\csname urlstyle\endcsname\relax
  \providecommand{\doi}[1]{doi: #1}\else
  \providecommand{\doi}{doi: \begingroup \urlstyle{rm}\Url}\fi

\bibitem[Athey and Imbens(2016)]{athey-2016-recursive-partitioning-hte}
Susan Athey and Guido Imbens.
\newblock Recursive partitioning for heterogeneous causal effects.
\newblock \emph{Proceedings of the National Academy of Sciences}, 113\penalty0
  (27):\penalty0 7353--7360, jul 2016.
\newblock \doi{10.1073/pnas.1510489113}.
\newblock \url{https://doi.org/10.1073%2Fpnas.1510489113}.

\bibitem[Bretz et~al.(2009)Bretz, Maurer, Brannath, and
  Posch]{bretz-2009-graphical}
Frank Bretz, Willi Maurer, Werner Brannath, and Martin Posch.
\newblock A graphical approach to sequentially rejective multiple test
  procedures.
\newblock \emph{Statistics in Medicine}, 28\penalty0 (4):\penalty0 586--604,
  2009.

\bibitem[Burman et~al.(2009)Burman, Sonesson, and
  Guilbaud]{burman-2009-recycling}
Carl-Fredrik Burman, Christian Sonesson, and Olivier Guilbaud.
\newblock A recycling framework for the construction of bonferroni-based
  multiple tests.
\newblock \emph{Statistics in Medicine}, 28\penalty0 (5):\penalty0 739--761,
  2009.

\bibitem[Chernozhukov et~al.(2018)Chernozhukov, Demirer, Duflo, and
  Fernandez-Val]{chernozhukov-2018-generic}
Victor Chernozhukov, Mert Demirer, Esther Duflo, and Ivan Fernandez-Val.
\newblock Generic machine learning inference on heterogeneous treatment effects
  in randomized experiments, with an application to immunization in india.
\newblock Technical report, National Bureau of Economic Research, 2018.

\bibitem[Cox(1975)]{cox-1975-splitting}
David~R. Cox.
\newblock A note on data-splitting for the evaluation of significance levels.
\newblock \emph{Biometrika}, 62\penalty0 (2):\penalty0 441--444, 1975.

\bibitem[Di~Francesco(2026)]{difrancesco-2026-aggregation-trees}
Riccardo Di~Francesco.
\newblock Aggregation trees.
\newblock \emph{Econometric Reviews}, 45\penalty0 (5):\penalty0 634--659, 2026.

\bibitem[Dmitrienko et~al.(2009)Dmitrienko, Tamhane, and
  Bretz]{dmitrienko-2009-multiple-testing}
Alex Dmitrienko, Ajit~C. Tamhane, and Frank Bretz.
\newblock \emph{Multiple testing problems in pharmaceutical statistics}.
\newblock CRC Press, January 2009.
\newblock ISBN 9781584889847.

\bibitem[Finner and Roters(2001)]{finner-2001-fdr}
Helmut Finner and Markus Roters.
\newblock On the false discovery rate and expected type i errors.
\newblock \emph{Biometrical Journal}, 43\penalty0 (8):\penalty0 985--1005,
  2001.

\bibitem[Foster et~al.(2011)Foster, Taylor, and
  Ruberg]{foster-2011-virtual-twins}
Jared~C. Foster, Jeremy M.~G. Taylor, and Stephen~J. Ruberg.
\newblock Subgroup identification from randomized clinical trial data.
\newblock \emph{Statistics in Medicine}, 30\penalty0 (24):\penalty0 2867--2880,
  2011.

\bibitem[Goeman and Solari(2010)]{goeman-2010-sequential-rejection}
Jelle~J. Goeman and Aldo Solari.
\newblock The sequential rejection principle of familywise error control.
\newblock \emph{The Annals of Statistics}, 38\penalty0 (6):\penalty0
  3782--3810, 2010.
\newblock \doi{10.1214/10-AOS829}.

\bibitem[Gutierrez and G{\'e}rardy(2017)]{gutierrez-2017-uplift}
Pierre Gutierrez and Jean-Yves G{\'e}rardy.
\newblock Causal inference and uplift modelling: A review of the literature.
\newblock In \emph{International conference on predictive applications and
  APIs}, pages 1--13. PMLR, 2017.

\bibitem[Huang et~al.(2025)Huang, Tang, and Kenney]{huang-2025-distilling-hte}
Melody Huang, Tiffany~M. Tang, and Ana~M. Kenney.
\newblock Distilling heterogeneous treatment effects: Stable subgroup
  estimation in causal inference.
\newblock 2025.

\bibitem[Maurer(1995)]{maurer-1995-fixed-sequence}
Willi Maurer.
\newblock Multiple comparisons in drug clinical trials and preclinical assays:
  a-priori ordered hypothesis.
\newblock \emph{Biomed. Chem.-Pharm. Ind.}, 6:\penalty0 3--18, 1995.

\bibitem[Meinshausen(2008)]{meinshausen-2008-hierarchical}
Nicolai Meinshausen.
\newblock Hierarchical testing of variable importance.
\newblock \emph{Biometrika}, 95\penalty0 (2):\penalty0 265--278, 2008.

\bibitem[Ondra et~al.(2016)Ondra, Dmitrienko, Friede, Graf, Miller, Stallard,
  and Posch]{ondra-2016-targeted-subgroups}
Thomas Ondra, Alex Dmitrienko, Tim Friede, Alexandra Graf, Frank Miller, Nigel
  Stallard, and Martin Posch.
\newblock Methods for identification and confirmation of targeted subgroups in
  clinical trials: a systematic review.
\newblock \emph{Journal of Biopharmaceutical Statistics}, 26\penalty0
  (1):\penalty0 99--119, 2016.

\bibitem[Powers et~al.(2018)Powers, Qian, Jung, Schuler, Shah, Hastie, and
  Tibshirani]{powers-2018-uplift}
Scott Powers, Junyang Qian, Kenneth Jung, Alejandro Schuler, Nigam~H. Shah,
  Trevor Hastie, and Robert Tibshirani.
\newblock Some methods for heterogeneous treatment effect estimation in high
  dimensions.
\newblock \emph{Statistics in medicine}, 37\penalty0 (11):\penalty0 1767--1787,
  2018.

\bibitem[Rubin et~al.(2006)Rubin, Dudoit, and van~der
  Laan]{rubin-2006-sample-splitting}
Daniel Rubin, Sandrine Dudoit, and Mark~J. van~der Laan.
\newblock A method to increase the power of multiple testing procedures through
  sample splitting.
\newblock Technical report, bepress, 2006.

\bibitem[Rubin(2004)]{rubin-2004-teaching-stats}
Donald~B. Rubin.
\newblock Teaching statistical inference for causal effects in experiments and
  observational studies.
\newblock \emph{Journal of Educational and Behavioral Statistics}, 29\penalty0
  (3):\penalty0 343--367, 2004.
\newblock ISSN 10769986, 19351054.
\newblock \url{http://www.jstor.org/stable/3701358}.

\bibitem[Shaffer(1986)]{shaffer-1986-modified}
Juliet~Popper Shaffer.
\newblock Modified sequentially rejective multiple test procedures.
\newblock \emph{Journal of the American Statistical Association}, 81\penalty0
  (395):\penalty0 826--831, 1986.

\bibitem[Wager and Athey(2018)]{wager-2018-uplift}
Stefan Wager and Susan Athey.
\newblock Estimation and inference of heterogeneous treatment effects using
  random forests.
\newblock \emph{Journal of the American Statistical Association}, 113\penalty0
  (523):\penalty0 1228--1242, 2018.

\bibitem[Wasserman and Roeder(2009)]{wasserman-2009-screen-and-clean}
Larry Wasserman and Kathryn Roeder.
\newblock High dimensional variable selection.
\newblock \emph{The Annals of Statistics}, 37\penalty0 (5A):\penalty0 2178,
  2009.

\bibitem[Wiens(2003)]{wiens-2003-fallback}
Brian~L. Wiens.
\newblock A fixed sequence bonferroni procedure for testing multiple endpoints.
\newblock \emph{Pharmaceutical Statistics: The Journal of Applied Statistics in
  the Pharmaceutical Industry}, 2\penalty0 (3):\penalty0 211--215, 2003.

\bibitem[Wiens and Dmitrienko(2005)]{wiens-2005-fallback-procedure}
Brian~L. Wiens and Alexei Dmitrienko.
\newblock The fallback procedure for evaluating a single family of hypotheses.
\newblock \emph{J Biopharm Stat}, 15\penalty0 (6):\penalty0 929--42, 2005.
\newblock \doi{10.1080/10543400500265660}.
\newblock \url{https://pubmed.ncbi.nlm.nih.gov/16279352/}.

\bibitem[Wu et~al.(2022)Wu, Tan, Li, Garrard, Obeng, Dimmery, Singh, Wang,
  Jiang, and Bakshy]{wu-2022-interpretable-personalized}
Han Wu, Sarah Tan, Weiwei Li, Mia Garrard, Adam Obeng, Drew Dimmery, Shaun
  Singh, Hanson Wang, Daniel Jiang, and Eytan Bakshy.
\newblock Interpretable personalized experimentation.
\newblock In \emph{Proceedings of the 28th ACM SIGKDD Conference on Knowledge
  Discovery and Data Mining}, pages 4173--4183, 2022.

\bibitem[Yadlowsky et~al.(2025)Yadlowsky, Fleming, Shah, Brunskill, and
  Wager]{yadlowsky-2025-rate}
Steve Yadlowsky, Scott Fleming, Nigam Shah, Emma Brunskill, and Stefan Wager.
\newblock Evaluating treatment prioritization rules via rank-weighted average
  treatment effects.
\newblock \emph{Journal of the American Statistical Association}, 120\penalty0
  (549):\penalty0 38--51, 2025.

\end{thebibliography}

\end{document}